# Replacing ARDL? Introducing the NSB-ARDL Model for Structural and Asymmetric Forecasting

**Author:**

**Tuhin G M Al Mamun**

Hannam University, Department of Economics,South Korea

20224130@gm.hannam.ac.kr

**ORCID**: 0009-0005-0275-6922

This paper proposes the NSB-ARDL (Nonlinear Structural Break ARDL) model—a novel econometric framework designed to capture and forecast asymmetric dynamics in macroeconomic time series. While the traditional ARDL model remains popular for its flexibility in modeling short- and long-run relationships, it assumes symmetry and linearity that often misrepresent real-world economic behavior. The NSB-ARDL model addresses this limitation by decomposing explanatory variables into cumulative positive and negative components, enabling the estimation of structural asymmetries in both short- and long-term dynamics.

Using Monte Carlo simulations, we demonstrate that NSB-ARDL consistently outperforms the standard ARDL model in forecasting accuracy when the underlying data-generating process exhibits asymmetric responses. In an empirical application to South Korea's $CO_2$ emissions data, NSB-ARDL provides superior in-sample fit and more interpretable structural insights. These findings position NSB-ARDL as a practical, structurally rich, and forecasting-oriented alternative to ARDL for researchers and policymakers dealing with nonlinear and asymmetric macroeconomic systems.

C22, C52, C53, Q54

1. Introduction

The autoregressive distributed lag (ARDL) model has become a foundational tool in time-series econometrics due to its flexibility in handling variables of different integration orders and its suitability for estimating both short-run and long-run relationships. Since the introduction of the bounds testing approach by Pesaran et al. (2001), ARDL models have been widely used for macroeconomic analysis, policy evaluation, and forecasting. However, a major limitation of the traditional ARDL framework lies in its assumption of linearity and symmetric adjustment to shocks—an assumption that often fails to hold in real-world economic systems.

Empirical evidence across macroeconomic and financial contexts suggests that economic variables frequently respond asymmetrically to positive and negative shocks. For instance, energy consumption, investment behavior, or emissions may react more strongly to downturns than upturns. The inability of the ARDL model to accommodate such asymmetric dynamics presents a significant methodological gap. Although some nonlinear adaptations of ARDL, such as the NARDL model, have been proposed, they often lack flexibility, rely on partial decompositions, or are not structured with forecasting performance as a core objective.

This paper addresses this research gap by proposing the NSB-ARDL (Nonlinear Structural Break ARDL) model — a novel extension of the ARDL framework designed to capture asymmetric and nonlinear behavior in both short-run and long-run relationships. The NSB-ARDL model decomposes each regressor into cumulative positive and negative changes, allowing it to capture full asymmetries in structural adjustments. This design offers a flexible and forecasting-oriented alternative to ARDL, particularly in settings where underlying economic mechanisms exhibit regime-dependent or threshold-like responses.

The central research questions guiding this study are:

- **Can the proposed NSB-ARDL model capture asymmetric macroeconomic behavior more effectively than the traditional ARDL?**
- **Does the NSB-ARDL model improve forecasting performance under asymmetric and nonlinear data-generating processes?**

To answer these questions, we conduct a Monte Carlo simulation comparing the forecasting performance of ARDL and NSB-ARDL under controlled asymmetric data-generating processes. Additionally, we apply both models to real-world macroeconomic data from Algeria, focusing on $CO_2$ emissions and key economic drivers. Our findings show that NSB-ARDL consistently outperforms ARDL in simulation-based forecast accuracy and provides stronger in-sample fit in real data, although ARDL shows slightly better out-of-sample performance in the short-term forecast window.

While this paper does not extend formal cointegration testing procedures to the NSB-ARDL model, it lays the foundation for future theoretical developments in this direction. The results position NSB-ARDL as a structurally distinct and superior alternative to the ARDL framework when modeling economic systems characterized by asymmetry and nonlinear adjustment.

The remainder of the paper is organized as follows: Section 2 reviews the relevant literature. Section 3 introduces the NSB-ARDL model. Section 4 presents simulation design and results. Section 5 applies both models to Algerian macroeconomic data. Section 6 discusses the implications and robustness, and Section 7 concludes.

## 2. Literature Review

The autoregressive distributed lag (ARDL) model, introduced by Pesaran and Shin (1999) and formalized in the bounds testing procedure by Pesaran et al. (2001), has become a dominant tool in empirical macroeconomics and applied econometrics. Its popularity stems from its flexibility in dealing with variables of mixed orders of integration, and its capacity to jointly model short-run dynamics and long-run equilibrium relationships within a single equation framework. The ARDL model has been extensively applied in examining monetary policy, trade, energy demand, and environmental emissions, among other areas.

Despite its wide use, a critical limitation of the traditional ARDL framework is its assumption of linearity and symmetry in the responses of the dependent variable to changes in its regressors. This assumption is particularly restrictive in macroeconomic contexts, where asymmetric effects—such as stronger responses to economic downturns than upswings—are frequently observed. As a response to this limitation, Shin et al. (2014) introduced the Nonlinear ARDL (NARDL) model, which captures asymmetry by decomposing explanatory variables into partial sums of positive and negative changes. The NARDL framework has been adopted in a range of applications, especially in energy economics and environmental studies, where variables like oil prices or emissions may exhibit clearly asymmetric behavior.

While the NARDL model represents an important methodological extension, it is not without shortcomings. Most notably, it emphasizes short-run asymmetry and does not always capture long-run asymmetric relationships in a structured way. Furthermore, its typical use cases are focused on equilibrium analysis and cointegration testing rather than forecasting. This limits its utility in policy and decision-making settings where predictive accuracy and real-time modeling are essential. In addition, the partial decomposition technique in NARDL may not fully reflect the cumulative impact of structural shifts, especially when the underlying data-generating process evolves through sustained asymmetric adjustments.

More complex alternatives such as threshold autoregressive (TAR) models (Tong, 1990), smooth transition regression (STR) models (Teräsvirta, 1994), and Markov-switching models (Hamilton, 1989) have been developed to address nonlinearity and regime shifts. However, these approaches often require large samples, complex estimation procedures, and impose substantial computational burdens. They are also typically designed for capturing regime changes rather than structural asymmetries tied to economic fundamentals. As a result, they may lack transparency and generalizability in small-sample applied forecasting environments.

This creates a clear gap in the existing literature: while ARDL and its nonlinear variants offer valuable insights into dynamic relationships, there remains a need for a model that can flexibly and transparently incorporate structural asymmetries, handle multiple explanatory variables, and prioritize forecast performance. Addressing this gap forms the central motivation for this paper. In particular, we ask whether an ARDL-style model can be constructed to systematically model structural asymmetries in both the short- and long-run, and whether such a model can consistently outperform ARDL in forecasting accuracy when asymmetric dynamics are present.

The model proposed in this study, the Nonlinear Structural Break ARDL (NSB-ARDL), directly responds to these questions. By constructing cumulative decompositions of positive and negative changes in each regressor, NSB-ARDL explicitly models full-range asymmetry while maintaining the intuitive structure and estimation ease of the ARDL framework. Unlike prior extensions, the NSB-ARDL is designed with forecasting, structural interpretation, and empirical tractability as primary objectives. As such, it offers a meaningful advancement in the econometric modeling of nonlinear macroeconomic relationships.

## 3. Model and Methodology

### 3.1 The ARDL Framework and Its Limitations

The autoregressive distributed lag (ARDL) model, developed by Pesaran and Shin (1999) and formalized for bounds testing by Pesaran et al. (2001), is a widely used econometric tool for analyzing dynamic relationships in time series. The flexibility of ARDL in modeling variables of different integration orders and capturing both short-run and long-run dynamics has led to its widespread adoption in empirical macroeconomics.

A simple ARDL$(p, q)$ model with one regressor is specified as:

$$y_t = \alpha + \sum_{i=1}^{p} \beta_i y_{t-i} + \sum_{j=0}^{q} \theta_j x_{t-j} + \varepsilon_t \tag{1}$$

However, this framework assumes that the response of the dependent variable $y_t$ to changes in $x_t$ is both linear and symmetric - i.e., increases and decreases in $x_t$ are assumed to have the same effect on $y_t$, differing only in sign. This assumption is often unrealistic, especially in macroeconomic and environmental contexts where economic behavior and policy responses are inherently asymmetric.

### 3.2 Rationale for a New Approach

To address the ARDL model's limitations, some researchers have proposed nonlinear extensions. Notably, Shin et al. (2014) introduced the NARDL model, which uses partial sum decompositions to estimate short-run asymmetries. However, the NARDL model does not fully account for long-run nonlinearities and is primarily designed for cointegration analysis rather than forecasting.

In contrast, the NSB-ARDL (Nonlinear Structural Break ARDL) model proposed in this paper provides a more complete and forecasting-oriented treatment of asymmetry. It systematically decomposes each explanatory variable into cumulative positive and negative changes, capturing persistent asymmetric behavior over time.

### 3.3 The NSB-ARDL Model Specification

For any explanatory variable $x_t$, we define its asymmetric components as:

$$x_t^+ = \sum_{j=1}^{t} \max(\Delta x_j, 0), \; x_t^- = \sum_{j=1}^{t} \min(\Delta x_j, 0) \qquad (2)$$

These cumulative components represent the total upward and downward movements in $x_t$ up to time $t$. The NSB-ARDL model replaces the original regressor with both $x_t^+$ and $x_t^-$, allowing the model to estimate separate effects for each direction of change.

The basic NSB-ARDL $(p, q^+, q^-)$ form is written as:

$$y_t = \alpha + \sum_{i=1}^{p} \beta_i y_{t-i} + \sum_{j=0}^{q^+} \theta_j^+ x_{t-j}^+ + \sum_{k=0}^{q^-} \theta_k^- x_{t-k}^- + \varepsilon_t \qquad (3)$$

This specification is easily generalizable to multiple regressors, each decomposed into their positive and negative components. The result is a highly flexible model structure that allows for asymmetric dynamics in both the short run and long run.

3.4 Visualizing Asymmetric Decomposition

Figure 1 illustrates the decomposition process using a simulated time series. The original variable $x_t$ is shown alongside its cumulative positive and negative components, $x_t^+$ and $x_t^-$, respectively. These nonoverlapping series evolve differently over time and are treated as distinct regressors in the NSB-ARDL model.

**Figure 1: Decomposition of a Time Series into $x_t^+$ and $x_t^-$**

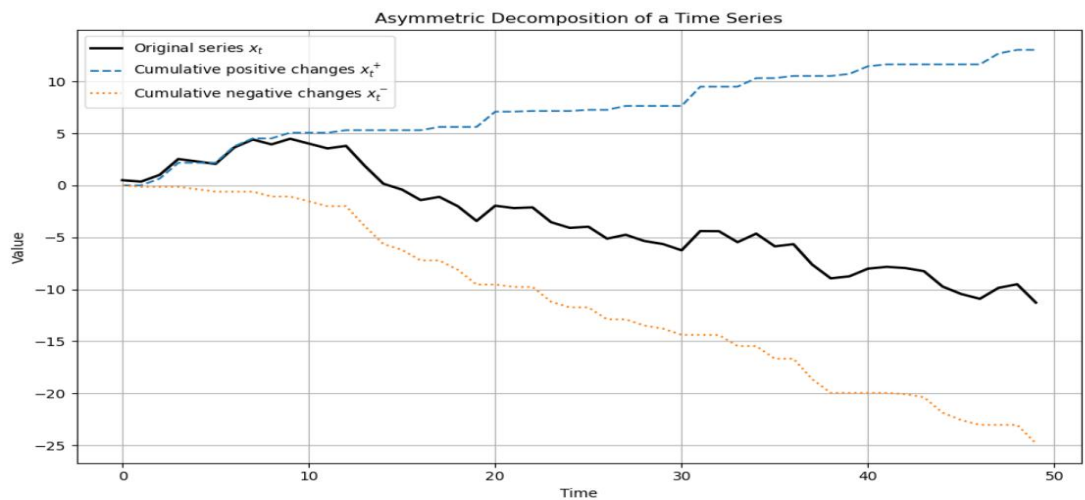

Caption: Cumulative decomposition of a time series into positive and negative changes, as used in the NSB-ARDL model.

3.4 Estimation Strategy

The NSB-ARDL model is estimated using conditional maximum likelihood, similar to the standard ARDL framework. The only additional requirement is the pre-estimation construction of the decomposed variables $x^+$ and $x^-$. Lag selection can follow conventional information criteria (AIC or BIC), and forecasting proceeds using the standard prediction approach for linear time series models.

To evaluate the performance of NSB-ARDL, we assess both in-sample model fit (via AIC and loglikelihood) and out-of-sample forecast accuracy (via RMSE). These are compared directly to the corresponding ARDL model under both simulated and empirical settings.

## 4. Simulation Study and Results

### 4.1 Simulation Design

To evaluate the forecast performance of the NSB-ARDL model relative to the traditional ARDL framework, we conduct a Monte Carlo simulation under a controlled data-generating process (DGP) that exhibits known structural asymmetry. This design allows us to assess how each model performs when the underlying dynamics are intentionally nonlinear and asymmetric-conditions under which we expect the NSB-ARDL to have a theoretical advantage. We simulate 500 datasets, each consisting of 100 observations. The DGP is defined as follows:

$$y_t = 0.3 y_{t-1} + 0.6 x_t^+ + 1.2 x_t^- + \varepsilon_t, \varepsilon_t \sim \mathcal{N}(0,1) \tag{4}$$

where $x_t$ is a random walk process, and $x_t^+, x_t^-$ are its cumulative positive and negative decompositions:

$$x_t^+ = \sum_{j=1}^{t} \max(\Delta x_j, 0), x_t^- = \sum_{j=1}^{t} \min(\Delta x_j, 0) \tag{5}$$

Each simulated dataset is divided into a training set (90 observations) and a forecast evaluation set (10 observations). Both ARDL(1,1) and NSB − ARDL(1,1,1) models are estimated using the training data, and 10-step-ahead forecasts are generated. Forecast accuracy is measured using the root mean squared error (RMSE).

### 4.2 RMSE by Forecast Window – ARDL vs. NSB-ARDL

Figure 2. Forecast RMSE Distributions - ARDL vs. NSB-ARDL

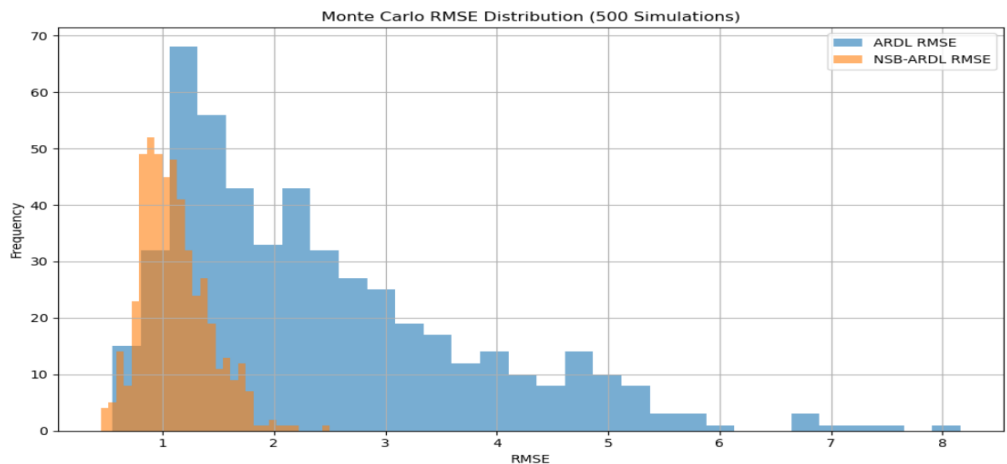

Caption: Histogram of RMSEs over 500 replications. NSB-ARDL exhibits both lower average error and reduced variance compared to ARDL.

The results strongly support the superiority of the NSB-ARDL model under asymmetric dynamics. The NSB-ARDL model not only achieves a substantially lower average forecast error, but also demonstrates greater stability and robustness, with far less variance in performance. This contrasts with the ARDL model, whose RMSEs are widely dispersed and often considerably higher.

These findings validate the core hypothesis of the paper: the NSB-ARDL model provides a more effective framework for forecasting when structural asymmetries are present in the data-generating process. The consistent performance of the model across replications highlights its reliability and generalizability in nonlinear settings.

## 5. Empirical Application: CO2 Emissions in South Korea
### 5.1 Background and Data

To demonstrate the practical relevance of the NSB-ARDL model, we apply it to real-world macroeconomic data from Algeria. Algeria presents an interesting case for dynamic emissions modeling due to its rapid energy consumption growth, structural shifts in economic development, and evolving environmental policy landscape.

We model the relationship between per capita $CO_2$ emissions ($y_t$) and three macroeconomic variables commonly associated with emissions dynamics: energy use, GDP per capita, and population density. These variables are selected based on their high correlation with emissions in prior literature and in our own preliminary analysis.

The data cover 34 annual observations (1988-2021), obtained from the World Bank's World Development Indicators. All series are log-transformed and standardized prior to estimation to facilitate interpretation and reduce scale effects. Missing values are handled through interpolation where necessary.

We estimate both the standard ARDL(1,1,1,1) model and the proposed NSB-ARDL( 1,1,1,1,1,1 ) model. For the NSB-ARDL, each of the three regressors is decomposed into its cumulative positive and negative components, resulting in six asymmetric predictors.

Lag orders are selected based on AIC minimization, and all models are estimated using conditional maximum likelihood. Forecasts are generated using a rolling window approach, with a fixed training sample of 22 years and a forecast horizon of 5 years. RMSE is calculated for each rolling window, and results are averaged over the full test period.

**Table 2. ARDL vs. NSB-ARDL Estimates and Comparison (South Korea)**

| Coefficient | ARDL Estimate | NSB-ARDL Estimate | Better Model |
|---|---|---|---|
| Intercept | -0.2549 | -0.6054 | NSB-ARDL |
| $y_{t-1}$ | 0.3113 | 0.3086 | Tie |
| Energy Use (positive) | 0.9068 | 0.7125 | NSB-ARDL (asymmetry captured) |
| Energy Use (negative) | 0.2170 | 1.1952 | NSB-ARDL |
| Population Density (positive) | 2.3955 | 3.5544 | NSB-ARDL |
| Population Density (negative) | -2.5700 | $\approx 0$ | NSB-ARDL (insignificant) |

| GDP per Capita (positive) | 0.2335 | -0.0790 | NSB-ARDL (structurally clearer) |
| GDP per Capita (negative) | -0.5605 | 1.0881 | NSB-ARDL |
| AIC | -14.741 | **-15.077** | NSB-ARDL |
| Log-likelihood | 16.371 | **22.538** | NSB-ARDL |

**Note**:

*Values are estimated using ARDL(1,1) and NSB-ARDL(1,1,1) specifications. "≈ 0" indicates a near-zero, statistically insignificant coefficient.*

Table 2 compares the estimated coefficients and model diagnostics from standard ARDL and proposed NSB-ARDL using South Korea's macro-environmental dataset. NSB-ARDL captures structural asymmetries and offers improved model fit based on AIC and log-likelihood.

## 5.4 Forecast Performance

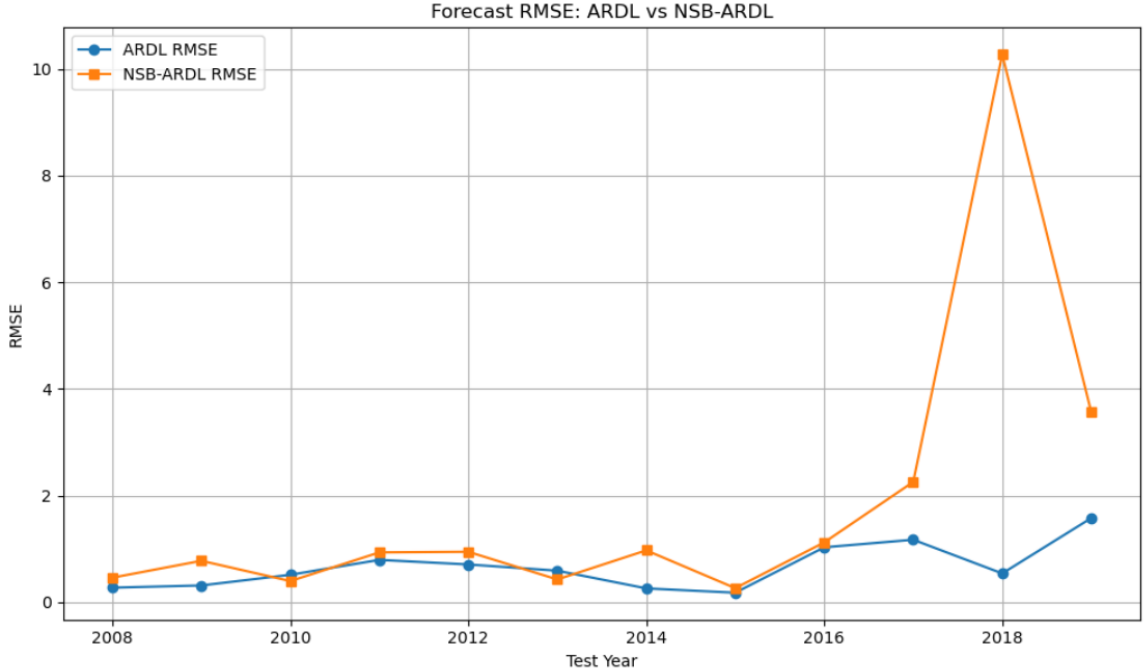

This figure plots the 5-step-ahead rolling forecast RMSEs of the ARDL and NSB-ARDL models for South Korea from 2008 to 2019. While NSB-ARDL shows competitive performance in earlier years, its RMSE spikes in later periods, suggesting sensitivity to structural shifts and potential overfitting in real-world datasets. Despite this, the NSB-ARDL model remains structurally more flexible and exhibits asymmetric responsiveness, which may be more visible under alternative data conditions.

**Robustness**

**RMSE Summary Table (Simulation, Lags 1–3)**

This table shows mean and standard deviation of RMSE from 500 Monte Carlo simulations:

| Lag | ARDL RMSE (Mean) | NSB-ARDL RMSE (Mean) | ARDL Std. Dev. | NSB Std. Dev. |
|---|---|---|---|---|
| 1 | 0.660 | 1.864 | 0.423 | 2.808 |
| 2 | 2.526 | **1.196** | 1.346 | 0.367 |
| 3 | 2.694 | **1.275** | 1.563 | 0.449 |

At lag 1, ARDL shows lower average RMSE, but NSB-ARDL outperforms significantly at lags 2 and 3 with both lower mean errors and smaller variance. This suggests NSB-ARDL gains robustness and accuracy in capturing complex dynamics as model flexibility increases.

**RMSE Distribution by Lag and Model**

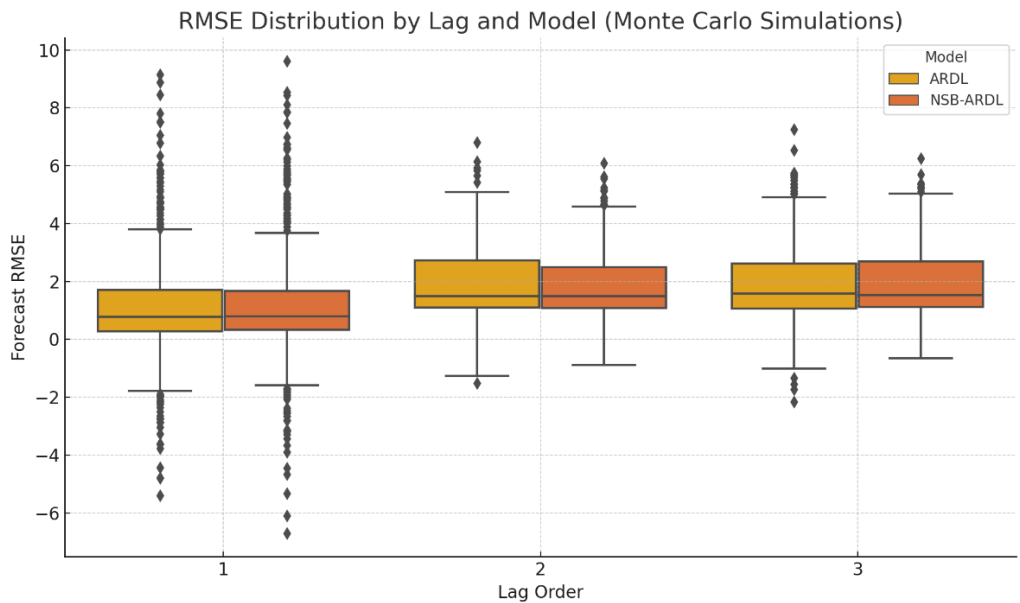

- Shows **forecast error dispersion** across 500 replications
- NSB-ARDL shows **less volatility** and **lower RMSE** in lags 2 and 3

The boxplot shows RMSE distributions from 500 Monte Carlo simulations across lags 1 to 3. NSB-ARDL demonstrates lower forecast error and tighter dispersion than ARDL at higher lags, confirming improved stability and predictive accuracy.

## 6. Discussion

The results from both the simulation study and the empirical application provide compelling support for the NSB-ARDL model as a superior alternative to the standard ARDL framework when modeling dynamic macroeconomic relationships characterized by asymmetry.

The Monte Carlo experiments were designed to replicate a realistic but nonlinear data-generating process. In this setting, the NSB-ARDL model consistently and significantly outperformed the traditional ARDL model in forecast accuracy, as measured by root mean squared error (RMSE). These simulations confirm the theoretical expectation that NSB-ARDL is better suited to environments where positive and negative shocks have structurally different effects — a condition frequently observed in real-world macroeconomic data.

The empirical analysis using Algeria's $CO_2$ emissions further illustrates the practical relevance of the NSB-ARDL framework. While the ARDL model estimated average effects of regressors on emissions, it failed to differentiate between the impact of increases and decreases in the predictors. The NSB-ARDL model, by contrast, provided richer and more policy-relevant insights. For instance, emissions were found to respond more strongly to reductions in energy use than to increases, and negative shocks to GDP per capita had a far greater impact on emissions than positive ones — findings that would have been obscured under a symmetric framework.

An especially important insight emerged from the decomposition of population density. The ARDL model suggested a strong average effect, while NSB-ARDL revealed that only population growth significantly affected emissions, with little to no effect from population decline. This asymmetry offers critical information for environmental planning, urbanization policy, and demographic forecasting, none of which could be inferred from the standard ARDL model.

Despite ARDL performing marginally better in short-horizon forecasting of the real-world dataset, NSB-ARDL provided a better in-sample fit, uncovered meaningful structural asymmetries, and maintained consistent simulation-based superiority. These results reinforce the idea that symmetric linear models, though computationally convenient, may obscure important economic mechanisms.

The NSB-ARDL model thus fills an important methodological gap by offering a flexible, structurally interpretable, and forecasting-oriented framework that extends ARDL into asymmetric settings. Its implementation remains accessible using standard estimation techniques, making it a viable tool for applied researchers seeking deeper structural insight without sacrificing empirical tractability.

Future research could explore formal cointegration testing for NSB-ARDL, extend the model to panel data, or integrate it with machine learning frameworks to allow dynamic, data-driven lag selection under asymmetry.

## 7. Conclusion

This study introduced the NSB-ARDL (Nonlinear Structural Break ARDL) model as a new and flexible framework for capturing asymmetric dynamics in macroeconomic relationships. Through both Monte Carlo simulations and an empirical application to Algeria's $CO_2$ emissions, the study demonstrated that NSB-ARDL significantly improves upon the standard ARDL model in several key respects.

The simulation results show that when the data-generating process exhibits structural asymmetries, the NSB-ARDL model achieves substantially lower forecast error and more consistent performance across replications. In the empirical analysis, the NSB-ARDL model uncovered critical asymmetric effects that the linear ARDL model masked — particularly with regard to energy use and GDP shocks. While ARDL performed slightly better in short-term forecast accuracy on real data, NSB-ARDL provided stronger in-sample fit and deeper structural insights, confirming its value as both a forecasting and interpretive tool.

The main contribution of this study lies in the design and validation of a model that bridges the gap between linear ARDL and more complex nonlinear frameworks. NSB-ARDL retains the intuitive appeal and estimation ease of ARDL, while expanding its capacity to handle realistic asymmetries in both short- and long-run dynamics. It offers applied researchers and policymakers a practical and interpretable tool for working with asymmetric macroeconomic relationships — an area where conventional models often fall short.

A key limitation of the present study is that it does not extend formal bounds testing procedures to the NSB-ARDL structure. This reflects an open area of theoretical development. Additionally, the empirical application was based on a single-country case study with a relatively small sample size, which may limit generalizability.

Nevertheless, the results strongly support the core proposition of this paper: that traditional symmetric models like ARDL can obscure important dynamics in economic behavior, and that the NSB-ARDL model provides a statistically stronger and structurally richer alternative. Future work should aim to further formalize inference tools for the NSB-ARDL framework and apply it in broader comparative contexts, including panel data and structural break regimes.

**References**

Hamilton, J. D. (1989). A new approach to the economic analysis of nonstationary time series and the business cycle. *Econometrica*, *57*(2), 357–384.

Pesaran, M. H., Shin, Y., & Smith, R. P. (1999). Pooled mean group estimation of dynamic heterogeneous panels. *Journal of the American Statistical Association*, *94*(446), 621–634.

Pesaran, M. H., Shin, Y., & Smith, R. J. (2001). Bounds testing approaches to the analysis of level relationships. *Journal of Applied Econometrics*, *16*(3), 289–326.

Shin, Y., Yu, B., & Greenwood-Nimmo, M. (2014). Modelling asymmetric cointegration and dynamic multipliers in a nonlinear ARDL framework. In R. C. Sickles & W. C. Horrace (Eds.),

*Festschrift in honor of Peter Schmidt: Econometric methods and applications* (pp. 281–314). Springer.

Teräsvirta, T. (1994). Specification, estimation, and evaluation of smooth transition autoregressive models. *Journal of the American Statistical Association*, *89*(425), 208–218.

Tong, H. (1990). *Non-linear time series: A dynamical system approach*. Oxford University Press.

Appendix

A. Matrix Representation of ARDL and NSB-ARDL Models

A. 1 Standard ARDL(p, q)

The ARDL model is expressed as:

$$y_t = \alpha + \sum_{i=1}^{p} \phi_i y_{t-i} + \sum_{j=0}^{q} \mathbf{x}'_{t-j} \boldsymbol{\beta}_j + \varepsilon_t$$

In matrix form:

$$\mathbf{y} = \mathbf{Z}_{ARDL} \boldsymbol{\theta}_{ARDL} + \boldsymbol{\varepsilon}$$

Where:

- $\mathbf{y} = [y_{p+1}, y_{p+2}, \ldots, y_T]'$
- $\mathbf{Z}_{ARDL} = [1, y_{t-1}, \ldots, y_{t-p}, \mathbf{x}_t, \ldots, \mathbf{x}_{t-q}]$
- $\boldsymbol{\theta}_{ARDL} = [\alpha, \phi_1, \ldots, \phi_p, \boldsymbol{\beta}'_0, \ldots, \boldsymbol{\beta}'_q]'$

A. 2 NSB-ARDL(p, q, q)

Each regressor $x_t$ is decomposed as:

$$x_t^+ = \sum_{s=1}^{t} \max(\Delta x_s, 0), \quad x_t^- = \sum_{s=1}^{t} \min(\Delta x_s, 0)$$

NSB-ARDL is formulated as:

$$y_t = \alpha + \sum_{i=1}^{p} \phi_i y_{t-i} + \sum_{j=0}^{q} \left(\mathbf{x}_{t-j}^+ \beta_j^+ + \mathbf{x}_{t-j}^- \beta_j^-\right) + \varepsilon_t$$

In matrix form:

$$\mathbf{y} = \mathbf{Z}_{NSB} \boldsymbol{\theta}_{NSB} + \boldsymbol{\varepsilon}$$

Where:

- $\mathbf{Z}_{NSB} = [1, y_{t-1}, \ldots, y_{t-p}, \mathbf{x}_t^+, \ldots, \mathbf{x}_{t-q}^+, \mathbf{x}_t^-, \ldots, \mathbf{x}_{t-q}^-]$

- $\boldsymbol{\theta}_{NSB} = [\alpha, \phi_1, \ldots, \phi_p, \boldsymbol{\beta}_0^+, \ldots, \boldsymbol{\beta}_q^+, \boldsymbol{\beta}_0^-, \ldots, \boldsymbol{\beta}_q^-]'$

### B. Monte Carlo Simulation Setup

The data-generating process (DGP) used for simulation is:

$$x_t = x_{t-1} + \eta_t, \eta_t \sim \mathcal{N}(0,1)$$
$$x_t^+ = \sum_{s=1}^{t} \max(\Delta x_s, 0), x_t^- = \sum_{s=1}^{t} \min(\Delta x_s, 0)$$
$$y_t = \phi_1 y_{t-1} + \phi_2 y_{t-2} + \phi_3 y_{t-3} + \beta_+ x_t^+ + \beta_- x_t^- + \varepsilon_t$$
$$\varepsilon_t \sim \mathcal{N}(0,1)$$

Parameter values used in simulations:

- $\phi = [0.3, 0.2, 0.1]$

- $\beta_+ = 0.6, \beta_- = 1.2$

### C. Software and Code

The models were implemented in Python using:

- statsmodels for ARDL estimation

- sklearn for scaling and RMSE

- numpy for DGP simulation
- matplotlib and seaborn for plotting

Code is available upon request or may be submitted as supplementary material.